\documentstyle[preprint,eqsecnum,aps]{revtex}

\begin{document}
\draft
\preprint{DOE/ER/40561-103-INT93-00-36; DOE/ER/40427-09-N93}
\title{The Off Shell $\rho - \omega$ Mixing in the QCD Sum Rules}
\author{
T.Hatsuda$^{(1)}$\thanks{Internet address: hatsuda@bethe.npl.washington.edu},
E.M.Henley$^{(1,2)}$\thanks{Internet address:
henley@alpher.npl.washington.edu},
Th.Meissner$^{(2)}$\thanks{Internet address: meissner@ben.npl.washington.edu}
and G.Krein$^{(3)}$\thanks{Internet address: gkrein@ift.uesp.ansp.br} }
\address{
$^{(1)}$ Physics Department, FM-15, University of Washington,
Seattle, WA 98195\\
$^{(2)}$ Institute for Nuclear Theory, HN-12, University of Washington,
Seattle, WA  98195\\
$^{(3)}$ Instituto de F\'{\i}sica Te\'orica, Rua Pamplona, 145, 01405-900
S\~ao Paulo-SP, Brazil\\ }

\date{\today}
\maketitle
\begin{abstract}
The $q^2$ dependence of the $\rho-\omega$ mixing amplitude
is analyzed with the use of
the QCD sum rules and the dispersion relation.
Going off shell the mixing decreases, changes sign at
$q^2 \simeq 0.4 m_{\rho}^2 > 0$ and is negative in the space like region.
Implications of this result to
the isospin breaking part of the nuclear force are discussed.
\end{abstract}
\pacs{ }

\section{Introduction}
\label{sec:s1}

Although charge independence and charge symmetry are respected approximately by
the strong interactions, these symmetries are both broken by the
electromagnetic interaction and the mass difference of the up and down quarks.
The symmetry-breaking remains of considerable interest both experimentally and
theoretically because it can be studied by perturbation theory.
At low
energies, the large scattering length of two nucleons in the $^1S_0$ state
makes it particularly suitable for investigations of both charge independence
and charge symmetry \cite{MILLER}.
At medium energies, polarization
experiments in n-p scattering have been of interest for studies of charge
symmetry \cite{AB,KNU}.
In nuclei, the mass difference between light-to-medium
conjugate nuclear pairs (with a neutron replaced by a proton) also shows charge
symmetry-breaking beyond that due to Coulomb forces; this is commonly referred
to as the Nolen-Schiffer or Okamoto-Nolen-Schiffer anomaly \cite{NS,SHL}.

Theoretically, charge-dependent forces have been cast into 4 classes \cite{HM}.
It is the class IV forces, proportional to $(\tau_1^{(3)}-\tau_2^{(3)})$
and $(\tau_1 \times \tau_2)^{(3)}$, where $\tau_i^j$ is the
jth component of the Pauli isospin matrix for particle $i$, which produce
charge-symmetry-breaking effects in the n-p system.
In meson-theoretic
investigations \cite{MTW}, these forces are due to (in addition to photon
exchange) pion exchange,
mixed $\rho -\omega$ and $\pi - \eta -\eta'$ exchanges,
combined
$\pi-\gamma$ and other complex exchanges.  The experimental
asymetries measured at $477 {\rm MeV}$ \cite{AB} and $183 {\rm MeV}$
\cite{KNU} are
particularly sensitive to pion-exchange and $\rho-\omega$ mixed exchange,
respectively.  Indeed, both experiments agree with theoretical predictions
based on meson theory \cite{MTW}.

At $183 {\rm MeV}$, the dominant contribution to the polarization asymmetry in
the n-p
elastic scattering cross section is found to arise
from $\rho-\omega$ mixing in the exchanged
mesons, as shown in fig.\ref{figure1}.
However, it has been pointed out by Goldman,
Henderson and Thomas \cite{GHT}
that the fit makes use of the on-mass-shell mixing
of the $\rho$ and $\omega$, as measured in $e^+e^-$ collisions, whereas it is
space-like values of the 4-momentum transfer ($q^2 <0$) that are required for
the fit to the n-p scattering data.  Furthermore, they point out that in a
model which uses free quark-antiquark intermediate states, there is
considerable
variation of the rho-omega mixing parameter with $q^2$; indeed, it changes
sign at $q^2 \sim -(400 {\rm MeV})^2$ and is small for negative (space-like)
values of $q^2$. This casts doubt
on the theoretical calculations \cite{MTW}.
More recently, Piekarewicz and Williams \cite{PW} have used an intermediate
$N\bar{N}$ to calculate the mixing and obtain qualitatively
similar conclusions as Goldman
et al.

The Nolen-Schiffer anomaly has also seen a revival of interest.  In a
meson-theoretic investigation, the type III forces proportional to
$(\tau_1^{(3)} + \tau_2^{(3)})$, in particular the $\rho-\omega$ mixing
contribution, is claimed to be an essential ingredient to explain the
anomaly \cite{BI}.
However, the on-mass-shell value of the $\rho-\omega$  mixing
is used also in this analysis.
The partial restoration of chiral symmetry in nuclei and the associated
change of the n-p mass difference in medium is another explanation of the
anomaly without recourse to the rho-omega mixing \cite{HK}.  The
effective theory of QCD \cite{HK} and  QCD sum rules in a nuclear
medium \cite{HHP}
seem to support this explanation.
Further extension of  the approach in ref.\cite{HHP} is also made by taking
into account the mixing in the vector channel \cite{SKB}.
Thus it is important to find out the relative importance of the
$\rho-\omega$  mixing among other explanations by analyzing
the $q^2$ variation of the mixing.

Of the various non-perturbative approaches in QCD,
QCD sum rules have been found
to be a powerful tool for analyzing the properties and decays of hadrons
\cite{SVZ0,REVIEW}.  They can give a plausible explanation of the
Nolen-Schiffer
anomaly \cite{HHP} as we mentioned above.
One of the earliest applications of the sum rules was actually
to a calculation of the on-mass shell $\rho-\omega$ mixing \cite{SVZ}.

In this paper, we follow the path forged by Shifman, Vainshtein and Zakharov
(SVZ) \cite{SVZ}, but we extend it to space-like values of $q^2$ which is
the relevant region of the charge-dependent nuclear forces.
Furthermore, we re-evaluate the various
semi-phenomenological constants
required
by the sum rules.  Whereas SVZ used a value of $\beta$ (see below) obtained
from
a ``speculation'' based on ``the experience with sum rules", or as a free
parameter, we obtain $\beta$ from the sum rules themselves.
In section 2, we will summarize the effect of the $\rho-\omega$  mixing
on the nuclear force and give an essential idea about the importance of the
$q^2$-dependence of the mixing.  In section 3, we will determine
the parameters which control the $q^2$ variation from
QCD sum rules.  The combined use of the Borel sum rules and the finite energy
sum rules together with the current knowledge of the quark mass difference
and the quark condensates give us an unambiguous determination of these
parameters without further assumptions.
We find that some of the outputs are different in magnitude and sign
from the SVZ results.
In section 4, the $q^2$ dependence of the $\rho$-$\omega$
mixing and also its effect
on the nuclear force are discussed by using the parameters determined in
section 3.  The effective mixing angle $\theta(q^2)$ changes sign
at $q^2 \simeq 0.4 m_{\rho}^2$ and
gets negative in the space like region, which
is qualitatively consistent with the previous analysis \cite{GHT,PW}; however
the effect is stronger in our case.
In coordinate space, this corresponds to a node of the charge-dependent
potential around $r \simeq 0.9{\rm fm}$ which suppresses the effect
of the $\rho-\omega$ mixing in the nuclear force.
Section 5 is devoted to a summary and concluding remarks.

\section{$\rho-\omega$ mixing in the space like region}
\label{sec:s2}

The Feynman graph for the class IV forces originating from $\rho-\omega$ mixing
is
shown in fig.\ref{figure1}.
The cross in fig.\ref{figure1} denotes the
$\rho-\omega$ mixing
$\theta(q^2)$ defined in terms of the mixed propagator
\begin{eqnarray}
\Pi_{\mu \nu}^{\rho \omega} (q^2) & = & i \int d^4 x e^{i qx} \langle {\rm T}
\rho_{\mu}(x)
\omega_{\nu}(0) \rangle_0 , \\  \nonumber
& = & (g_{\mu \nu} - {q_{\mu} q_{\nu}\over q^2})
       { \theta(q^2) \over (q^2 - m_{\rho}^2+i
\epsilon)(q^2-m_{\omega}^2+i\epsilon)}  ,
\label{eq:21}
\end{eqnarray}
where we have neglected the width of the resonances for simplicity.

At $q^2=m_{\omega}^2$,  $\theta(q^2)$ reduces to the mixing matrix
$\langle \rho \mid H_{SB} \mid \omega \rangle $ measured in the $e^+e^-
\rightarrow \pi^+ \pi^-$
data \cite{CB}
\begin{eqnarray}
\theta(q^2=m_{\omega}^2) & = &  \langle \rho \mid H_{SB} \mid \omega \rangle
\nonumber \\
 & = & (-4520 \pm 600) \ \ {\rm MeV}^2 .
\label{eq:22}
\end{eqnarray}
Here $H_{SB}$ is a part of the Hamiltonian which breaks isospin symmetry and we
have used the covariant
definition of the mixing matrix
$\langle \rho_{\mu} \mid H_{SB} \mid \omega_{\nu} \rangle = - (g_{\mu \nu} -
q_{\mu} q_{\nu}/q^2)
\langle \rho \mid H_{SB} \mid \omega \rangle$ following  ref.\cite{CS}.

$H_{SB}$ has both QED and QCD origins.  One of the QED effects
to eq.(2.1); from $\rho \rightarrow \gamma \rightarrow
\omega$; can be explicitly calculated and gives a small
and positive contribution \cite{FEY}
\begin{eqnarray}
\theta_{\rho \rightarrow  \gamma
\rightarrow \omega}(m_{\omega}^2) =
{e^2 \over  m_{\rho}^2 g_{\rho} g_{\omega}} \simeq 610 \ \ {\rm MeV}^2 ,
\label{eq:23}
\end{eqnarray}
where $e /g_{\rho}$ and $e/g_{\omega}$ denote,respectively, the
electromagnetic coupling of $\rho$ and $\omega$ with $\gamma$.
We have  assumed $g_{\omega}\simeq 3g_{\rho}$,as
obtained from the flavor SU(3) symmetry.
Thus the main part of the observed
 mixing comes from the quark mass difference
\begin{eqnarray}
m_{d} - m_{u} \simeq 0.28 (m_d+ m_u) ,
\label{eq:24}
\end{eqnarray}
where $(m_{d}+m_u)/2=(7 \pm 2){\rm MeV}$ at a $1 {\rm GeV}$  QCD scale
(\cite{GLR} and ref.therein).
Note also that,  once one takes into account the $\gamma$ exchange between
nucleons
with proper nucleon electro-magnetic form factors (fig.\ref{figure2}),
one has to subtract the $\rho \rightarrow \gamma \rightarrow \omega $
effect from fig.\ref{figure1} to avoid  double counting. (The QED mixing is
hidden
in the form factors in fig.\ref{figure2}.)

For isospin mixing in the nuclear force, we are concerned with
space-like momenta
$q^2 < 0$,
which is far from the on shell point $q^2=m_{\rho,\omega}^2$.
Thus it is crucial to study the $q^2$
variation of $\theta(q^2)$ to determine whether one can naively utilize
the on-shell mixing value, eq.(2.2), in the space like region \cite{GHT}.
In the next section, we will analyze the $q^2$ variation of $\rho-\omega$
mixing on the basis of the QCD sum rules
in which
the physical quantities are related to the condensates in the QCD vacuum.  Our
approach
has less ambiguities compared to  other approaches based on
phenomenological models \cite{GHT,PW}.
Before entering into details, let us discuss the primary origin of the $q^2$
variation to
demonstrate the essential idea in a model-independent way.

We start with the unsubtracted dispersion relation for $\Pi_{\mu \nu}^{\rho
\omega}(q^2)$:
\begin{eqnarray}
 {\rm Re} \Pi_{\mu \mu}^{\rho \omega}(q^2)
 = {{\rm P} \over \pi} \int_0^{\infty} {{\rm Im} \Pi_{\mu \mu}^{\rho \omega}(s)
\over
 s-q^2} ds.
\label{eq:25}
\end{eqnarray}
If  $q^2 $ is not too large and in the space-like region, the main contribution
to ${\rm Re}
\Pi_{\mu \mu}^{\rho \omega}$ comes from the lowest poles in the imaginary part
of $\Pi_{\mu \mu}^{\rho \omega}$, i.e. the $\rho$ and
$\omega$ resonances.  It is important to note that
for the mixing, $\Pi_{\mu \mu}^{\rho \omega}$, unlike the case for
$\Pi_{\mu \mu}^{\rho \rho}$,  the integral of the continuum contribution such
as that from an
$N \bar{N}$ loop is finite \cite{PW}.
If photon exchange is included in the loop, this is no longer the case;
however, we find this contribution is to be small numerically ($\approx 2\%$),
so that it can be neglected.
Thus
we do not need any subtraction
in the right hand side of eq.(2.4)
as far as the strong interaction is concerned.  In the  pole
approximation,  ${\rm Im} \Pi_{\mu \mu}^{\rho \omega}(s)$ is easily evaluated
by saturating the
intermediate states by $\rho$- and $\omega$-mesons:
\begin{eqnarray}
{1 \over 3} {\rm Im} \Pi_{\mu \mu}^{\rho \omega}(s) & = &
\pi \langle 0| \rho_{\mu} |\rho \rangle \langle \rho | \omega_{\mu} | 0 \rangle
\delta(s-m_{\rho}^2)
 + \pi \langle 0| \rho_{\mu} |\omega \rangle \langle \omega | \omega_{\mu} | 0
\rangle \delta(s-m_{\omega}^2)
\nonumber \\
& \equiv & \pi F_{\rho} \delta(s-m_{\rho}^2) -  \pi F_{\omega}
\delta(s-m_{\omega}^2) .
\label{eq:26}
\end{eqnarray}
Since $m_{\rho} \neq m_{\omega}$, the residues at the $\rho$
and  $\omega$-poles do not have
to be equal
\begin{eqnarray}
F_{\rho} - F_{\omega} = {\cal O}(m_d-m_u) .
\label{eq:27}
\end{eqnarray}
Substituting (2.5) into (2.4), one obtains
\begin{eqnarray}
{1 \over 3}{\rm Re} \Pi_{\mu \mu}^{\rho \omega}(q^2) & = & -
[ {F_{\rho} \over q^2-m_{\rho}^2} - {F_{\omega} \over q^2 - m_{\omega}^2}] ,
\nonumber  \\
& = &  { {\delta m^2 (F_{\rho}+F_{\omega})/2 - (q^2 - m^2)(F_{\rho} -
F_{\omega})}
 \over (q^2-m_{\rho}^2) (q^2-m_{\omega}^2)} ,
\label{eq:28}
\end{eqnarray}
where $m^2 \equiv (m_{\rho}^2+m_{\omega}^2)/2$
and $\delta m^2 \equiv m_{\omega}^2-m_{\rho}^2
 = {\cal O}(m_d-m_u)$.

Thus we finally get
\begin{eqnarray}
\theta(q^2)
 = \theta(m^2) [1+ \lambda ({q^2 \over m^2} -1)],
\label{eq:29}
\end{eqnarray}
with $\theta(m^2) = \delta m^2 (F_{\rho}+F_{\omega})/2 $ and
 $\lambda = -(F_{\rho}-F_{\omega})m^2 / \theta(m^2)$.
The first term of eq.(2.8) is
nothing but the usual on-shell mixing matrix
$\langle \rho | H_{SB} | \omega \rangle$, while the second term gives
a $q^2$ variation due to $F_{\rho} \neq F_{\omega}$.  Since both the first and
second terms are
proportional to $m_d-m_u$ and are of the same order, there is no a priori
reason to neglect the second
term.
However, in most phenomenological applications \cite{MILLER,MTW,BI} of
$\rho-\omega$ mixing to the charge symmetry breaking nuclear
force, $F_{\rho}=F_{\omega}$ has been  assumed without any justification.

The $q^2$ variation of
$\theta(q^2)$ is an inevitable consequence of the isospin symmetry breaking and
we have not used any specific models up to this stage.
However, the sign and the magnitude of $F_{\rho} - F_{\omega}$ (or $\lambda$ )
in eq.(2.8)
depends on the QCD dynamics. In QCD, we have to start with the quark correlator
instead of the
hadronic correlator $\Pi_{\mu \nu}^{\rho \omega}$:
\begin{eqnarray}
\Pi_{\mu \nu} (q^2) = i \int d^4 x e^{iqx}
\langle {\rm T } J_{\mu}^{\rho}(x) J_{\nu}^{\omega}(0) \rangle_0,
\label{eq:210}
\end{eqnarray}
where
\begin{eqnarray}
J_{\mu}^{\rho} = (\bar{u} \gamma_{\mu} u -\bar{d}\gamma_{\mu} d)/2 \ \ \ \ \
J_{\mu}^{\omega} = (\bar{u} \gamma_{\mu} u +\bar{d}\gamma_{\mu} d)/6.
\label{eq:2100}
\end{eqnarray}
These  vector
currents couple to $\rho$ and $\omega$ as well as to
the higher resonances ($\rho', \omega', \cdot \cdot \cdot$) and the continuum.
Thus the imaginary part of $\Pi_{\mu \nu}$
 is written as  the sum of all the mixed hadron correlators
\begin{eqnarray}
{\rm Im}\Pi_{\mu \mu}(s) = A_0 {\rm Im}\Pi_{\mu \mu}^{\rho \omega}(s)
 + A_1 {\rm Im} \Pi_{\mu \mu}^{\rho'\omega'}(s) + \cdot \cdot \cdot \ \ ,
\label{eq:211}
\end{eqnarray}
where $A_n (n=0,1,2, \cdot \cdot \cdot )$ denotes the overlap of the quark
current with the physical
hadrons. For example
\begin{eqnarray}
A_0 = {m_{\rho}^2 m_{\omega}^2 \over g_{\rho} g_{\omega}},
\label{eq:212}
\end{eqnarray}
where $g_{\rho,\omega}$ is defined by
$\langle 0 | J_{\mu}^{\rho,\omega} | \rho (\omega) \rangle =
(m_{\rho,\omega}^2/g_{\rho,\omega})
\epsilon_{\mu} $ with $g_{\omega}\simeq 3g_{\rho}$ as
given by flavor SU(3) and $g_{\rho}^2/4\pi \simeq 2.4$.

By using the dispersion relation for $\Pi_{\mu \mu}$ and the
OPE for ${\rm Re }\Pi_{\mu \mu}(q^2)$ at
$q^2 \rightarrow - \infty$, one can extract the resonance parameters
such as the
mixing matrices for $\rho-\omega$ and for higher resonances
in the r.h.s. of eq.(2.10).
The first attempt at such an analysis was
made by Shifman, Vainshtein and Zakharov \cite{SVZ}.  However,
one of the mixing parameters $\beta$, which is related to $\lambda$ as
$\lambda=\beta +1$,  was not determined well because of the poor $e^+e^-$ data
available at that time and of
the large uncertainty of the current quark masses.
 In the next section,  armed with the updated information
of these parameters (eqs. (2.2) and (2.3)), we will
determine the crucial quantity $\beta$ within the framework of the  QCD sum
rules.
The combined use of the finite energy sum rule and the Borel sum rule is a key
ingredient there.
{}From the analysis,
we find that $\lambda$ is indeed positive and, including all uncertainties
in the parameters, is found to lie in an interval:
\begin{eqnarray}
\lambda = \beta + 1 \in [1.43,1.85]
\label{eq:213}
\end{eqnarray}
$\beta$ is close to,
but generally larger than the value assumed in ref.\cite{SVZ}, 0.5.
It follows that
$\theta(q^2)$ has significant $q^2$ variation
in the space like region; indeed,  it vanishes at
\begin{eqnarray}
q^2 = m^2 {\lambda -1 \over \lambda} .
\label{eq:214}
\end{eqnarray}
Thus the nuclear force is significantly affected by the $q^2$ variation of
$\theta(q^2)$.  This can be  illustrated by examining the coordinate space
charge-symmetry-breaking nuclear force in the static limit which is related to
 the Fourier transform of  eq. (2.1):
\begin{eqnarray}
V_{NN}^{\rho \omega}(r) = - {g_{\rho N} g_{\omega N} \over 4 \pi}  {\theta(m^2)
\over 2 m} [1- {2 \lambda \over mr}] e^{-mr},
\label{eq:215}
\end{eqnarray}
where $g_{\rho (\omega) N}$ denotes the vector coupling constant of the nucleon
and vector mesons. (We did not consider the form factor for the coupling
constants,  because we are
interested in the longer range part of $V_{NN}^{\rho \omega}$.)
When $\beta=0$, eq.
(2.11) reduces to the potential given by Coon and Barret \cite{CB}.
The potential is strongly attenuated by the presence of $\beta$ and
becomes zero at
\begin{eqnarray}
r = {2 \lambda \over m} \sim 0.9\ \ {\rm fm} \ \ \ \ \  ,
\label{eq:216}
\end{eqnarray}
which is the region of interest for symmetry-breaking effect.

\section{$\rho-\omega$ mixing in the QCD sum rules}
\label{sec:s3}

\subsection{Operator Product Expansion (OPE)}
\label{sec:s31}

For the QCD sum rule determination of the $\rho - \omega$ mixing,
one should start with the current correlator \cite{SVZ}
given in eqs. (\ref{eq:210},\ref{eq:2100}).
Because the currents $J^\rho _\mu$ and $J^\omega _\nu$ are conserved,
$\Pi_{\mu\nu}$ has to be of transversal structure:
\begin{eqnarray}
\Pi _{\mu\nu} (q^2) = (q_\mu q_\nu - g_{\mu\nu} q^2 ) \Pi (q^2)
\label{eq:SVZD}
\end{eqnarray}
Following ref. \cite{SVZ}, we
will analyze sum rules obtained by the OPE for $\Pi(q^2)$.
In the OPE series, we keep power corrections up to order $6$.
Furthermore, in each term we take contributions which are
of ${\cal O} (m_q)$ ($q= u \; {\rm or} \; d$)
or of ${\cal O} (\alpha)$, and any terms
${\cal O}(m_q ^2) \, , \, {\cal O} (\alpha^2)$ and
${\cal O} (m_q \alpha)$ are neglected.
We will check the validity of these assumptions later on.
Therefore we are faced with the  diagrams in fig.\ref{figure3},
which contribute to
$\Pi (Q^2)$ ($Q^2 = - q^2$) as follows \cite{SVZ}:

\noindent
(1){\em Free Part} (fig.3a):
The leading order term proportional to $ \ln Q^2$ vanishes, while the next term
\begin{eqnarray}
\Pi_{(1)} = {3\over {2 \pi^2}} {{ m_d ^2 - m_u ^2}\over {12 Q^2}},
\label{eq:OPE1}
\end{eqnarray}
is of ${\cal O} (m_q ^2)$ and will be neglected.
For the same reason, the one-gluon exchange
diagram (fig.3b)
is of higher order in $m_q$.

\noindent
(2){\em Irreducible $1-\gamma$ exchange} (fig.3c):
Because of the different coupling of the $\gamma$ to the $u$ and $d$ quarks,
the one-$\gamma$ exchange in fig.3c contributes as
\begin{eqnarray}
\Pi_{(2)} = - {\alpha \over{ 16\pi^3}} {1\over {12}} \ln Q^2 .
\label{eq:OPE2}
\end{eqnarray}

\noindent
(3){\em Reducible $1-\gamma$ exchange} (fig.3d):
As has been proposed in ref.\cite{SVZ}, it is convenient to separate
the reducible $\gamma$ exchange diagram from the whole analysis from the very
beginning. This can be done by neglecting fig.3d in the OPE side and
subtracting $\rho\rightarrow\gamma\rightarrow\omega$ contribution
(\ref{eq:23}) from the phenomenological side.

\noindent
(4){\em Gluon Condensate} $\alpha_s \langle G^2 \rangle_0$ (fig.3e) :
 It vanishes for the same reason as (1).

\noindent
(5) {\em Quark condensate} $m_q \langle \bar{q} q \rangle_0 $ (fig.3f):
\begin{eqnarray}
\Pi_{(5)} = {2\over {12 Q^4}} [m _u \langle \bar{u} u \rangle_0
 - m_d \langle \bar{d} d \rangle_0 ] \ .
\label{eq:OPE5}
\end{eqnarray}

\noindent
(6) {\em Mixed Condensate} (fig.3g):
As has been shown in refs. \cite{SVZ,LNT}, even in case of only one quark
flavor, these condensates contribute at least in ${\cal O} (m_q ^2)$
to the vector
current and are therefore not present in our analysis.

\noindent
(7) {\em 4-quark condensate} (fig.3h-k):
There are contributions from gluon as well as $\gamma$ exchanges.
Typical diagrams are shown in figs.3h-k.

(a) (fig.3h)
\begin{eqnarray}
\Pi_{(7a)} (Q^2) = - {{2 \pi \alpha_s} \over {12 Q^6}}
\langle
 ( \bar{u} \gamma_\alpha \gamma_5 \lambda^a u)^2 -
 ( \bar{d} \gamma_\alpha \gamma_5 \lambda^a d)^2
\rangle_0 \ \ .
\label{eq:OPE7a}
\end{eqnarray}

(b) (fig.3i)
\begin{eqnarray}
\Pi_{(7b)} (Q^2) = - {{4 \pi \alpha_s} \over {9 \cdot 12 Q^6}}
\langle
 ( \bar{u} \gamma_\alpha \lambda^a u)^2 -
 ( \bar{d} \gamma_\alpha \lambda^a d)^2
\rangle_0  \ \ .
\label{eq:OPE7b}
\end{eqnarray}

(c) (fig.3j)
\begin{eqnarray}
\Pi_{(7c)} (Q^2) = - {{8 \pi \alpha} \over {12 Q^6}}
\langle
{4 \over 9}( \bar{u} \gamma_\alpha \gamma_5  u)^2 -
{1 \over 9}( \bar{d} \gamma_\alpha \gamma_5  d)^2
\rangle_0\ \ .
\label{eq:OPE7c}
\end{eqnarray}

(d) (fig.3k)
\begin{eqnarray}
\Pi_{(7d)} (Q^2) = - {{16 \pi \alpha} \over {9 \cdot 12 Q^6}}
\langle
{4 \over 9}( \bar{u} \gamma_\alpha  u)^2 -
{1 \over 9}( \bar{d} \gamma_\alpha  d)^2
\rangle_0 \ \ .
\label{eq:OPE7d}
\end{eqnarray}

\subsection{Symmetry-breaking parameters and vacuum condensates}
\label{sec:s32}

There are essentially two  quantities which control the isospin-symmetry
breaking in OPE:  $m_{d}/m_{u}$ and $\langle \bar{d}d \rangle_0/\langle
\bar{u}u \rangle_0$.
Although they are related in principle in QCD, they appear as
independent parameters in the QCD sum rules.
As we will see below,  the dimension 4 operator
in the OPE is controlled by the quark mass ratio while
the dimension 6 operator depends on the ratio of the condensates.
Because the dimension 4 operator is the dominant term in the OPE, the
$\rho-\omega$ mixing is essentially determined by the quark mass
ratio and depends weakly on the ratio of the condensates.

The quark mass ratio is determined by the analysis of the mass
 splitting  of mesons and barons
 using the chiral perturbation theory (\cite{GLR} and ref.therein):
\begin{eqnarray}
{{m_d - m_u} \over {m_d + m_u}} = 0.28 \pm 0.03 \ \ \ .
\end{eqnarray}
  The isospin breaking of the quark condensate has been analyzed by several
 methods;  the chiral
perturbation theory \cite{GL},  the QCD sum rules for scalar and pseudoscalar
mesons
\cite{NARI,DD,NARL}, effective models of QCD incorporating the
 dynamical breaking of chiral symmetry \cite{PRS,NJL} and  the QCD sum rules
for baryons \cite{ADI}.
 \begin{eqnarray}
 \gamma = {{\langle \bar{d} d \rangle_0} \over {\langle \bar{u} u \rangle_0}}
-1
 = \left\{ \begin{array}{ll}
           - (6 \sim 10) 10^{-3}  & \mbox{ \cite{GL}} \\
           - (10 \pm 3) 10^{-3}  &  \mbox{ \cite{NARI}} \\
           - (7 \sim 9) 10^{-3}   &   \mbox{\cite{PRS,NJL}} \\
           -(2 \pm 1) 10^{-3}  &  \mbox{\cite{ADI}}
    \end{array} \right. \ \ \ \ .
 \label{eq:r}
\end{eqnarray}
Here we have assumed $\langle \bar{s}s \rangle_0/\langle \bar{u}u \rangle_0 -1
 = -(15 \sim  30) \%$ to get the first number in the above.
The last number is obtained solely by using the baryon mass splittings in the
QCD sum rule and is quite different from the other
 determinations.

Besides the symmetry-breaking parameters, we need to fix the
 value of the four-quark condensate.  The vacuum saturation
 hypothesis gives a rough estimate of its magnitude;
it gives a satisfactory description of the $\rho$-meson
 properties \cite{SVZ0}.  However,  there are recent
 arguments which favor a value larger by a factor $2$ or more
 than that adopted in \cite{SVZ0}  (see e.g.
 \cite{REVIEW,GBP} and the references therein).
 Since this condensate is not determined well in the sum rule,
 we will take the following two typical numbers and
 carry out the Borel analysis using them:
 \begin{eqnarray}
\alpha_s \langle \bar{q} q \rangle^2_0 =
 \left\{   \begin{array}{ll}
  1.81 \cdot 10^{-4} {\rm GeV}^6  &  \mbox{
           vacuum saturation \cite{SVZ0}} \\
  3.81 \cdot 10^{-4} {\rm GeV}^6 & \mbox{ \cite{GBP}}
 \end{array} \right. \ \ \ .
\label{eq:vcon}
\end{eqnarray}

The definition of $\gamma$ in eq.(\ref{eq:r}) and the vacuum saturation
hypothesis allow us to
 rewrite the isospin breaking in the four-quark condensate as
 \begin{eqnarray}
\langle \bar{u} u \rangle ^2_0 - \langle \bar{d} d \rangle ^2_0
= -2\gamma \langle \bar{q} q \rangle ^2_0 \ \ ,
\end{eqnarray}
\begin{eqnarray}
\alpha \langle \bar{u} u \rangle ^2_0 = \alpha  \langle \bar{d} d \rangle ^2_0
=
\alpha \langle \bar{q} q \rangle ^2_0  \ \ .
\end{eqnarray}

Using the above symmetry-breaking parameters and the vacuum condensates
together with the Gell-Mann-Oakes-Renner relation \cite{GOR}
\begin{eqnarray}
(m_u + m_d) \langle \bar{u} u + \bar{d} d \rangle_0 = -2 f_\pi ^2 m_\pi ^2 ,
\end{eqnarray}
eq. (\ref{eq:OPE5}) and the sum of (\ref{eq:OPE7a})-(\ref{eq:OPE7d})
 are written as
\begin{eqnarray}
\Pi_{(5)} (Q^2) = {1\over{12Q^4}}
{{m_d - m_u} \over {m_d + m_u}}
2 f_\pi ^2 m_\pi ^2
\left [ 1 +
{{m_d + m_u} \over {m_d - m_u}}
{\gamma \over {2 + \gamma}} \right ] \ \ ,
\end{eqnarray}
and
\begin{eqnarray}
\Pi_{(7)} (Q^2) =
- {{224}\over{81}} {2 \over { 12 Q^6}} \pi \cdot
\left [ \alpha_s \langle \bar{q} q \rangle^2_0 \right ] \cdot \left [
 -\gamma + {\alpha \over {8 \alpha_s (\mu^2)}} \right ]\ \ .
\end{eqnarray}

Summing up all contributions, we obtain
\begin{eqnarray}
12 \Pi (Q^2) = - c_0 \ln Q^2 +
{{c_1}\over {Q^2}} +
{{c_2}\over {Q^4}} +
2{{c_3}\over {Q^6}}     \ \ ,
\label{eq:OPE}
\end{eqnarray}
with
\begin{eqnarray}
c_0 & = & {\alpha \over{ 16 \pi^3}}
\label{eq:c0} \\
c_1 & = & {3\over {2 \pi^2}} (m_d ^2 - m_u ^2) \sim 0
\label{eq:c1}\\
c_2 & = &
{{m_d - m_u} \over {m_d + m_u}}
2 f_\pi ^2 m_\pi ^2
\left [ 1 +
{{m_d + m_u} \over {m_d - m_u}}
{\gamma \over {2 + \gamma}} \right ]
\label{eq:c2} \\
c_3 & = &
- {{224}\over{81}}  \pi \cdot
\left [ \alpha_s \langle \bar{q} q \rangle^2_0 \right ] \cdot \left [
 - \gamma+ {\alpha \over {8 \alpha_s (\mu^2)}} \right ]
\label{eq:c3}.
\end{eqnarray}
For the renormalization point $\mu^2$, we take a typical scale
 of the Borel mass $1 {\rm GeV}^2$, so that
\begin{eqnarray}
\alpha_s = \alpha_s (1 {\rm GeV}^2) \sim 0.5 \; \; .
\end{eqnarray}
It is obvious from eqs. (\ref{eq:c2}, \ref{eq:c3}) that the
 $m_d/m_u$ controls the magnitude of the dim. 4 matrix element
 and $\langle \bar{d}d \rangle_0/\langle \bar{u}u \rangle_0$ controls
 the dim. 6 matrix element.

The aim of our sum rule analysis is to determine the
parameter $\beta$ (or $\lambda$); we will use the experimental
$\rho-\omega$ mixing at $q^2=m_{\omega}^2$ as an input.
 For later use, let's define here the dimensionless mixing
of hadronic origin as
\begin{eqnarray}
\xi = {{-12}\over {g_\rho g_\omega}} {{\theta(m^2) -
\theta_{\rho\rightarrow\gamma\rightarrow\omega} }\over {m^2}} = (11.3 \pm 1.3)
\cdot
10^{-4} \ .
\label{eq:xiex}
\end{eqnarray}

To take into account the large uncertainties of $\gamma$ and
$\alpha_s \langle \bar{q} q \rangle_0^2$, we perform our analysis for
the  4 different
sets of input parameters specified in table\ref{table1}
with the corresponding
OPE coefficients $c_i$ ($i=0\dots3$) shown in table\ref{table2}.
Borel stability analyses will be made for these four different sets
in section \ref{sec:s35}.

\subsection{Spectral Function and Sum Rule}
\label{sec:s33}

The general expression for the spectral function, i.e. the imaginary part of
$\Pi_{\mu\nu} (q^2)$ is given in eq.(\ref{eq:211}). As has been pointed out
in refs.\cite{SVZ,REVIEW}, it is absolutely necessary to keep at least one
higher resonance ($\rho^\prime$, $\omega^\prime$)
in (\ref{eq:211}) in order to obtain stability in the QCD sum rule. We will
treat the related
mass ${m^\prime}^2 = {1\over 2} ( m_{\rho^\prime} ^2 + m_{\omega^\prime} ^2 )$
as well as the corresponding coupling constants $g_{\rho^\prime}$ and
$g_{\omega^\prime} $ as effective parameters to
be determined from the Borel stability analysis. Furthermore,
in order to account
for the electromagnetic logarithm in eq.(\ref{eq:OPE2}) in
the space-like region, we have to introduce a continuum threshold $s_0$.
The corresponding part in the spectral function is `suppressed' by the
electromagnetic coupling $\alpha$ and indeed it will turn out that the
influence of this term on the sum rule analysis is very small.
Assuming sharp resonances for $\rho$,$\omega$,$\rho^{\prime}$ and
$\omega^{\prime}$ as well as a step function for the continuum one can write,
analogous to eq.(\ref{eq:26}):
 \begin{eqnarray}
\nonumber
12{\rm Im} \Pi (s)  & = &
\pi f_\rho \delta (s - m_\rho ^2 )  -
\pi f_\omega  \delta (s - m_\omega ^2 )
\\ \nonumber
& + &
\pi f_{\rho^\prime}  \delta (s - m_{\rho^\prime} ^2 )  -
\pi f_{\omega^\prime}  \delta (s - m_{\omega^\prime} ^2 )
\\
& + &  {\alpha \over {16 \pi^2}} \theta (s - s_0 ).
\label{eq:SVZspec}
\end{eqnarray}
The general form of the Borel sum rule reads \cite{SVZ}:
\begin{eqnarray}
{1\over {\pi M^2}} \int_0 ^ \infty d s e^{- s/M^2} {\rm Im} \Pi (s) =
\tilde{\Pi}_B (M^2),
\end{eqnarray}
where the Borel transform $\tilde{\Pi}_B (M^2)$ is defined by:
\begin{eqnarray}
\tilde{\Pi}_B (M^2) =
\lim_{\begin{array}{c} Q^2,n \rightarrow \infty \\
M^2 = Q^2 / n \; {\rm fixed} \end{array}}
{1 \over {(n-1)!}} Q^{2n} \left ( {{d^2}\over {dQ^2}} \right )^n \Pi (Q^2).
\end{eqnarray}
Let us define the following parameters
\begin{eqnarray}
\nonumber
\xi & = & {{\delta m^2}\over {m^4}}
{{f_\rho + f_\omega}\over 2} \ \ , \\
{\xi^\prime} & = & {{\delta {m^\prime} ^2}\over {{m^\prime} ^4}}
{{f_{\rho^\prime} + f_{\omega^\prime}}\over 2}   ,
\end{eqnarray}
\begin{eqnarray}
\nonumber
\beta = -(f_\rho- f_\omega ) m^2 \cdot \left [ {\delta m ^2}
{{f_\rho + f_\omega}\over 2}  \right ]^{-1} , \\
{\beta^\prime} = -(f_{\rho^\prime} - f_{\omega^\prime}) {m^\prime} ^2
\cdot \left [ {\delta {m^\prime} ^2}
{{f_{\rho^\prime} + f_{\omega^\prime}}\over 2}  \right ]^{-1}.
 \label{eq:beta}
\end{eqnarray}
$\xi$ and $\xi'$ are directly related to the on-mass-shell mixing
 in the $\rho-\omega$ and $\rho'-\omega'$ channels, respectively.
 $\beta$ and ${\beta^\prime}$ are related to the isospin-symmetry breaking
 at the resonance poles and control the $q^2$ dependence of the
 $\rho-\omega$ and $\rho'-\omega'$ mixing.
The relation of $\beta$ and $\lambda$ defined in section 2 is easily obtained
by comparing
 (\ref{eq:211}), (\ref{eq:212}), (\ref{eq:26}) with
 (\ref{eq:SVZD}) and (\ref{eq:SVZspec}):
\begin{eqnarray}
\lambda =  \beta + 1\ \ .
\end{eqnarray}

By using the above definitions,
one obtains from eq.(\ref{eq:OPE}) up to ${\cal O} (\delta m ^2)$ and ${\cal O}
(\delta
{m^\prime} ^2 )$ the Borel sum rule (BSR):
\begin{eqnarray}
\nonumber
& & \xi {{m^2}\over{M^2}} \left ( {{m^2}\over {M^2}}  - \beta \right )
e^{ - m^2 /M^2} +
{\xi^\prime} {{{m^\prime} ^2}\over{M^2}} \left ( {{{m^\prime} ^2}\over {M^2}} -
{\beta^\prime} \right )
e^{ - {m^\prime} ^2 /M^2} +
{\alpha \over {16\pi^3}} e^{ - s_0 /M^2}\\
& = &
c_0 +
c_1 \left ( {{m^2}\over {M^2}} \right ) +
c_2 \left ( {{m^2}\over {M^2}} \right )^2 +
c_3 \left ( {{m^2}\over {M^2}} \right )^3\ \ ,
\label{eq:BSR}
\end{eqnarray}
where $c_0$-$c_3$ are in units of $10^{-4}$.

Starting from this expression, one can carry out the Borel analysis (stability
analysis with respect to $M^2$) or analysis of the finite energy sum rule
(expansion in $1/M^2$ on both sides).
Let us summarize our strategy before entering
into the details.
On the right hand side of eq.(\ref{eq:BSR}),
we have 7 resonance parameters: $\beta$, ${\beta^\prime}$, $\xi$, $\xi'$,
$m$, $m'$ and $s_0$.  Among them, $m$ is the average mass of
$\rho$ and $\omega$, and $\xi$ is directly related to the measured
$\rho-\omega$ mixing, so they are known inputs.
$m'$ should be chosen to be a number close
to $m_{\rho'} = 1390 {\rm MeV}$ or
$m_{\omega'} = 1450 {\rm MeV}$; thus it is not  arbitrary.
The sum rule depends on $s_0$  very weakly because of the
suppression by $\alpha$; thus the choice of $s_0$ does not
materially affect the result.
Hence we are left with
essentially three unknown parameters $\beta$, ${\beta^\prime}$ and $\xi'$
to be determined by the sum rule.

In the finite energy sum rule, by expanding eq.(\ref{eq:BSR})
in powers of $1/M^2$,
we get three independent equations.
They are enough to solve for
the three unknowns.
One can also change $m'$ and $s_0$ slightly
to check the sensitivity of the result to that change.
One of the problems of the FESR when compared to the Borel sum rule
is that it is generally sensitive to the
parameters of the higher resonances.
Therefore, after getting a rough idea for the magnitude and sign of the
three
unknowns, we will move on to the Borel sum rule.

By taking a derivative of eq.(\ref{eq:BSR})
with respect to $M^2$, one generates another independent
sum rule.
To carry out the stability analysis for each unknown
parameter ($\beta$, ${\beta^\prime}$ and $\xi'$), we need one more
constraint, and we take a first sum rule of the
FESR to supply this constraint.
The first sum rule is known to be a local duality relation and is
least sensitive to the higher resonances.
By this procedure, we can draw independent Borel curves for each
unknown and carry out the stability analysis by varying
$m'$ and $s_0$, i.e. $m'$ and $s_0$ are chosen such that
$\beta$, ${\beta^\prime}$ and $\xi'$ are least sensitive to the variation
of the Borel mass $M^2$.
In this way, we can completely
determine the resonance parameters within the QCD sum rules.
Our procedure is different from the original SVZ analysis \cite{SVZ}
where $\beta$ was assumed to be 1/2 as an input and also poor
experimental data of $\xi$  was used.

\subsection{Finite Energy Sum Rule (FESR)}
\label{sec:s34}

 The FESR can be formally obtained by expanding the BSR
eq.(\ref{eq:BSR}) in powers of $1\over {M^2}$ and comparing the
coefficients of each power
on both sides. In doing so one obtains three FESR:
\begin{eqnarray}
 - \beta \xi-
{\beta^\prime} {\xi^\prime} {{{m^\prime}^2}\over{m^2}} - c_0 {{s_0}\over{m^2}}
& = & c_1
\label{eq:FESR1}
\\
\xi + \xi\beta + {\xi^\prime} {{{m^\prime}^4}\over{m^4}} +
{\xi^\prime}{\beta^\prime}
{{{m^\prime}^4}\over{m^4}}
+ {1\over 2} c_0 \left ( {{s_0}\over{m^2}} \right )^2 & = & c_2
\label{eq:FESR2}
\\
- \xi -
{1\over 2} \xi\beta - {\xi^\prime} {{{m^\prime}^6}\over{m^6}} - {1\over 2}
{\xi^\prime}{\beta^\prime}
{{{m^\prime}^6}\over{m^6}}
- {1\over 6} c_0 \left ( {{s_0}\over{m^2}} \right )^3 & = & c_3
\label{eq:FESR3}
\end{eqnarray}
Using various values for ${m^\prime} ^2$ and $s_0$ as input parameters we can
therefore solve the equations (\ref{eq:FESR1})-(\ref{eq:FESR3})
which gives us the 3 unknown
parameters $\beta$, ${\xi^\prime}$ and ${\beta^\prime}$.
Table\ref{table3} shows their values for the
OPE-set\ I. As we can see, the result is sensitive to the choice of the
resonance mass ${m^\prime} ^2$,
whereas the influence of the electromagnetically suppressed threshold $s_0$ is
small.

It is important to notice that $\beta\xi$ and ${\beta^\prime}{\xi^\prime}$ have
to be opposite
in sign due to the first sum rule of
 FESR (\ref{eq:FESR1}) or equivalently the local duality relation,
\begin{eqnarray}
\int_0 ^\infty ds {\rm Im} \Pi (s) = {\cal O} (\alpha) \approx 0\ \ ,
\label{eq:IMSR}
\end{eqnarray}
 This is the reason
 why one should keep the resonance explicitly \cite{SVZ},  otherwise
the term $\beta\xi$ cannot be canceled.

\subsection{Borel Analysis}
\label{sec:s35}

Finally we perform the Borel stability analysis of (\ref{eq:BSR}). Again we
take
for $\xi$ the experimental input (\ref{eq:xiex}). ${m^\prime} ^2$ and $s_0$ are
used
as variable  parameters which are fixed in order to obtain maximal
stability of the Borel curves.
In order to choose a stability window for $M^2$, we demand that, for each
parameter set (I-IV), the contribution of the $6$th order power correction
in the OPE is less than $25 \% $ of the $4$th order one.
This gives a lower limit $M_{min} ^2$ for the Borel mass $M^2$.

As we mentioned before, we adopt the first sum rule of the FESR
(\ref{eq:FESR1}) or equivalently the local duality relation
(\ref{eq:IMSR}) as an additional constraint. Also, we
generate another sum rule  from (\ref{eq:BSR}) by
operating with
$\left [ {1\over{M^4}} \left ( - {\partial\over{\partial
{(1/M^2)}}} \right ) M^2 \right ] $
on both sides.
In doing so one can uniquely calculate the $3$ unknown parameters $\beta$,
${\xi^\prime}$ and ${\beta^\prime}$ as function of $M^2$.
\begin{eqnarray}
\beta (M^2) & = & {1\over \xi}
{{Z_2 (M^2) +
Z_1 (M^2) \left ( 1 - {{{m^\prime}^2}\over{m^2}}\right ) } \over
{W_1 (M^2) \left ( {{{m^\prime}^2}\over{m^2}}- 1 \right ) - W_2 (M^2)} } \\
\label{eq:betaborel}
{\xi^\prime} (M^2) & = &  \left ( {{M^2}\over{m^2}} \right ) ^2
e^{{m^\prime}^2/M^2}
\left [ \beta (M^2) \xi W_1 (M^2) + Z_1 (M^2) \right ] \\
\label{eq:xip}
{\beta^\prime} (M^2) & = & {1\over {{\xi^\prime} (M^2)}} \left [ - \beta (M^2)
\xi
{{m^2}\over{{m^\prime}^2}} - c_0 {{s_0}\over{{m^\prime}^2}} - c_1
{{m^2}\over{{m^\prime}^2}}
\right ] \\
\label{eq:betap}
\end{eqnarray}
with:
\begin{eqnarray}
\nonumber
Z_1 (M^2) & = &
c_0 \left [ 1- e^{-s_0/M^2} - {{s_0}\over{M^2}} e^{-{m^\prime}^2/M^2} \right ]
+
c_1 {{m^2}\over{M^2}} \left ( 1 -
e^{-{m^\prime}^2/ M^2}
\right )
\\ \nonumber & - &
\xi
\left ( {{m^2}\over{M^2}} \right )^2
e^{-m^2 / M^2}
+ c_2
\left ( {{m^2}\over{M^2}} \right )^2  +
c_3
\left ( {{m^2}\over{M^2}} \right )^3 \\
\nonumber
W_1 (M^2) & = &
{{m^2}\over{M^2}} \left (
e^{-m^2 / M^2} -
e^{-{m^\prime}^2 / M^2} \right ) \\ \nonumber
Z_2 (M^2) & = & {1\over{M^4}} \left [
- {\partial\over{\partial
{\left (1/M^2\right )}}} \left ( M^2 Z_1 (M^2) \right ) \right ] \\
W_2 (M^2) & = & {1\over{M^4}} \left [
- {\partial\over{\partial
{\left (1/M^2\right )}}} \left ( M^2 W_1 (M^2) \right ) \right ]
\end{eqnarray}

The general procedure for determining ${m^{\prime}}^2$ and $s_0$ is now the
following:
For each of the parameter sets I-IV from table\ref{table1} we first take
the corresponding central value (i.e. without error bars) of the
OPE coefficients $c_i$ ($i=1\dots 3$) from table\ref{table2} and calculate
the Borel curves $\beta (M^2)$, $\xi^\prime (M^2)$ and $\beta^\prime
(M^2)$ due to eqs.
(\ref{eq:betaborel}), (\ref{eq:xip}) and (\ref{eq:betap}), respectively,
for various values of
${m^\prime}^2$ and $s_0$. The final result is the one with the least
dependence on $M^2$ within the Borel window $ M^2 \ge {M_{min}}^2$
with ${M_{min}}^2$ determined above.
The results for $\beta$ and $\xi^\prime$ using parameter set I
are shown in figs. \ref{figure4} and
\ref{figure5}, respectively. One recognizes a wide stable region,
which even can be extended up to $M^2 \approx 10 {\rm GeV}^2$.
The maximal stability occurs for ${m^\prime} ^2 = 1.6 {\rm GeV}^2$,
where $s_0$ can be increased up to about $2.4 {\rm GeV}^2$ without any
significant change.
This insensitivity of the result to the variation of the continuum threshold
are found for all other parameter sets as well and reflects the fact, that
the dependence on
$s_0$ is suppressed by the electromagnetic coupling $\alpha$.
With ${m^\prime}^2$ and $s_0$ obtained in this way, we now consider
the change in the Borel curves for $\beta$, $\xi^\prime$ and $\beta^\prime$
for the upper and
lower limits of the $c_i$ from table\ref{table2} and extract from this the
upper and lower limits for $\beta$, $\xi^\prime$ and $\beta^\prime$.
For $\beta$ (set I) the limits are shown in fig.\ref{figure6}.
The final values are given in table\ref{table4}.
It turns out that the variation of $\beta$, $\xi^\prime$ and $\beta^\prime$
is completely determined by the variation of the coefficient $c_2$, which
itself is based on the uncertainty of the quark mass difference
${{m_d -m_u}\over {m_d + m_u}} = 0.28 \pm 0.03$. The uncertainty of
$\gamma$, which creeps into the coefficient $c_3$ of the dimension $6$
condensate can be neglected.
It should be noted that a full readjustment of ${m^\prime}^2$ and $s_0$
for the upper and lower limits of the $c_i$'s from table\ref{table2}
which are obtained
by performing a new stability analysis for each case would result in
much bigger error bars for $\beta$, $\xi^\prime$ and $\beta^\prime$.

Similar considerations hold for the parameter sets III and IV.
In case of set II we find that the Borel curves are rather unstable
in the window $M^2 \ge {M_{min}}^2 = 1.5 {\rm GeV}^2$
(figs.\ref{figure7} and \ref{figure8}), especially the one for $\xi^\prime$,
and therefore this uncertainty, which is bigger than the variation due
to the $c_i$'s, has to be included in the corresponding error bars of
table\ref{table4}.

Finally we have convinced ourselves
that in none of these cases does the result change noticeably if one either
switches off the electromagnetic interaction completely, i.e. puts
$\alpha$ equal
to zero, or includes the ${\cal O} ({m_q}^2)$ diagram (fig.3a) with
the corresponding coefficient $c_1$ in the Borel analysis;
this analysis justifies
the neglect of these terms in higher orders.

We also want to stress that the value of $\beta$ obtained by this method
differs significantly from that used in ref. \cite{SVZ},
$\beta = 0.5$, which was based on a $SU(3)$ symmetry argument and served
as input parameter. Also the individual signs of ${\xi^\prime}$
and ${\beta^\prime}$ come
out different than in \cite{SVZ}, which arises from the fact, that
in addition to $\beta$, a different experimental value of $\xi$ was
used there.

\section{$\theta(q^2)$ and the $N-N$ potential}
\label{sec:s4}

In the previous section, we have determined the parameter
$\beta$ (=$\lambda-1$) which controls the $q^2$-dependence
of the $\rho-\omega$ mixing parameter $\theta(q^2)$.  The result summarized
in table\ref{table4} together with the relation between
$\lambda$ and $\theta(q^2)$ (eq.(\ref{eq:29})) immediately leads
to fig.\ref{figure9}.
The region inside the shaded band of fig.\ref{figure9} is favored
by our QCD sum rule analyses.

An important observation is that
$\lambda$ is always greater than 1 as long as $\beta$ is positive.
This feature leads to the conclusion that $\theta(q^2)$ already changes sign
for {\em positive} $q^2$ and is always negative
for $q^2 < 0$, as can be seen in fig.\ref{figure9}.
Although
the considerable variation of $\theta(q^2)$ obtained here
is qualitatively similar to previous results
based on different models, our $\theta(q^2)$ has a
stronger $q^2$ dependence than that of others.

In the quark-loop model ($\rho \rightarrow q \bar{q} \rightarrow
\omega$) with momentum cutoff, Goldman
et al. \cite{GHT},  $\theta(q^2)$ changes sign at a space like
momentum ($q^2 \sim - m_{\rho}^2/2$), therefore the
variation of $\theta(q^2)$ is more moderate than ours.
The nucleon-loop model ($\rho \rightarrow N \bar{N} \rightarrow
\omega$) with dimensional regularization \cite{PW} predicts
a sign change of $\theta(q^2)$ at $q^2=0$, which still
corresponds to a more moderate variation than our result.

One should notice here that there is a crucial assumption
with no theoretical justification in  both quark-loop
and nucleon-loop models:
the effect of the isospin breaking other than the QED effect
is solely attributed to the mass difference between
$u$ and $d$ constituent quarks (in the quark-loop model)
or to the mass difference between the proton and the neutron
(in the nucleon-loop model).
There is no a priori reason, however, to neglect the
isospin breaking in the coupling constants of the vector mesons
with the constituent quarks or the nucleons, which
generates an extra effect to the $\rho$-$\omega$ mixing
of ${\cal O}(m_d-m_u)$; nor is there justification for neglecting nuclear
resonances. e.g. $\Delta$, $N^*$.
Thus
it is fair to say that refs.\cite{GHT,PW} take into account only
a part of the total contribution due to the
current-quark mass difference.
We do not suffer from this deficiency, since
all the isospin-breaking effects are automatically taken
into account in the OPE of the current correlator.

Let us also comment on the unsubtracted
dispersion relation for $\Pi_{\mu \mu}(q^2)$ which
we have used to extract $\theta(q^2)$ from
${\rm Im} \Pi_{\mu \mu}(q^2)$.
As can be seen from the OPE in section 3, $\Pi_{\mu \mu}(q^2)$
approaches $m_d^2 - m_u^2$ as $q^2 \rightarrow - \infty$
if we switch off the small QED effect.  Therefore, the
dispersion integral is convergent and there is no need of
a subtraction to extract  ${\rm Re}\Pi_{\mu \mu}(q^2)$
for the entire $q^2$ region from the spectral function
obtained from the QCD sum rule.
This is the reason why we could  {\em predict}
the $q^2$ dependence of  the mixing.
If a subtraction were necessary, then extra experimental information
would have been required to fix the subtraction constant.
\footnote{When we  determine  the phenomenological parameters
in the spectral function
${\rm Im}\Pi_{\mu \mu}(s)$ of the QCD sum rule,
the subtraction is irrelevant since the Borel sum rule is used.}

The $q^2$ variation of $\theta (q^2)$ is of primary interest for
the isospin-breaking $N-N$
interaction shown in fig.\ref{figure1}.
The $N-N$ potential,
including the effect of $\theta(q^2)$, is given by
eq.(2.16).  To compare the potential $V_{NN}^{\rho \omega}(r)$
with and without
the $q^2$-dependence, we have shown it
in fig.\ref{figure10} for $\lambda=\beta+1$ (the dotted line)
and $\lambda=0$ the (the full line),
where we have used $\lambda = 0.78$ from set I (c.f.table\ref{table4}) as a
typical example.
The long range
Yukawa potential due to the $\rho-\omega$ mixing
(the first term in eq.(2.16)) is strongly suppressed
by the $q^2$ dependence of the mixing (second term in eq.(2.16)).
As a result,
in the coordinate space the potential changes sign
at $r = 2 \lambda /m =  0.9 {\rm fm}$.
The large difference between the solid and the dashed curves of fig.
\ref{figure10} at short
distances is not really relevant, since the short range
part of the potential $V_{NN}^{\rho \omega}(r)$
is screened by the strong central repulsion
of the nuclear force and also is reduced by the
the $\rho-N-N$ and $\omega-N-N$ form factors.

Several remarks are in order here on the $N-N$ potential due to
the isospin breaking:

\begin{enumerate}
\item In the conventional application of the
$\rho-\omega$ mixing to the $N-N$ potential, only the
mixing of the $\rho-\omega$ propagators is taken into account,
as is shown in fig.\ref{figure1}.
However, there is another
source of isospin-breaking  due to the
isospin-breaking $\rho (\omega)-N-N$ coupling
(c.f.fig.\ref{figure11}).
The magnitude of this symmetry breaking is unknown.
Furthermore this effect does not contribute to the polarization difference in
$n-p$ scattering, but does contribute to the Nolen-Schiffer anomaly.
\item The QCD sum rule can give us the mixing in
the $\rho-\omega$ channel and that in the
$\rho'-\omega'$ channel separately.
Therefore we
can study the effect of
$\rho'-\omega'$ mixing to  the $N-N$ force.
{}From the QCD sum rule we have extracted $\beta^\prime \approx -2.5 \dots
-3.5$,
so that the corresponding $\lambda^\prime = \beta^\prime + 1$ is clearly
negative and therefore the effect goes into the other direction, i.e. the
$\rho^\prime - \omega^\prime$ mixing angle $\theta^\prime (q^2)$ increases in
the space-like region.
The effect of $\theta^\prime (q^2)$ ($\rho^\prime - \omega^\prime$ mixing)
to the nuclear force will appear only at short distances compared to
$\theta (q^2)$ ($\rho - \omega$ mixing).
Thus it is not so relevant to the low energy $N-N$ scattering.
Since we do not have any information of the
$\rho^\prime (\omega^\prime) - N - N$ coupling constants, we cannot make any
quantitative estimate to the effect of $\theta^\prime (q^2)$ at the
present stage.
In the one boson exchange approach to the nuclear force the higher resonance
contributions are not taken into account explicitly.
\item In our approach we have taken a sharp resonance for the $\rho$ meson
spectral function and therefore completely neglected the width of the
$\rho$ meson.
However we believe that including the $\rho$ meson width does not change
our result significantly, because only the integral over the spectral
function is important in the sum rule and furthermore the crucial quantity
$\lambda = \beta +1 >1$ does not depend so much on the details of the
calculation of $\beta$.
\end{enumerate}

\section{Summary and Conclusions}
\label{sec:5}

QCD sum rules have been shown to be a powerful technique for studying meson
and baryon properties.
In this paper we have used these sum rules together with the usual unsubtracted
dispersion relation to study $\rho$-$\omega$ mixing both on- and off- mass
shell.
We find a rapid variation of the mixing parameter $\theta (q^2)$ with $q^2$.
Thus, although we fit the value of $\theta$ at $q^2 = {m_\rho} ^2 \approx
{m_\omega} ^2$, its sign has already changed at $q^2 = 0$.
Since it is the space-like value of $q^2$ that play a role in the contribution
of $\rho$-$\omega$ mixing to the nuclear force, our results have important
implications for fitting experimental results.
For example, the on-shell value of the $\rho$-$\omega$ mixing parameter,
$\theta (m^2)$,
has been used to fit the asymmetry observed in polarized $n$-$p$
scattering \cite{MILLER,AB,KNU,MTW}. Indeed, at $\approx 200 {\rm MeV}$,
this (assumed on-mass-shell)
mixing gives the dominant contribution to the observed asymmetry.
Our result, like those obtained previously by other methods
\cite{GHT,PW}, thus place a large question mark on our understanding
of the observed charge asymmetry.
In addition $\rho$-$\omega$ mixing contributes to the difference of the $p$-$p$
and $n$-$n$ scattering lengths and to the energy differences of mirror nuclei.
Indeed, there has been a revival of interest in the contribution of
$\rho$-$\omega$ mixing to the Nolen-Schiffer anomaly \cite{NS,SHL}
and our result also puts some of these explanations
\cite{BI,SKB} into questions.

\acknowledgements

This work has been supported in parts by the Department of Energy
(T.H.,E.H.,T.M.) and the Alexander von Humboldt-Stiftung
(Feodor-Lynen Program) (T.M.).
G.K.wishes to thank the Nuclear Theory Group of the University of Washington
for financial support.

\begin{figure}
\caption{Class IV nuclear force from the $\rho$-$\omega$ mixing}
\label{figure1}
\end{figure}

\begin{figure}
\caption{$\gamma$-exchange between two nucleons with electromagnetic form
factors}
\label{figure2}
\end{figure}

\begin{figure}
\caption{Diagrams in the OPE}
\label{figure3}
\end{figure}

\begin{figure}
\caption{Borel curve for $\beta$ with 3 different values of $({m^\prime}^2 \; ,
\; s_0)$
(Parameter-Set I)}
\label{figure4}
\end{figure}

\begin{figure}
\caption{Borel curve for ${\xi^\prime}$ with 3 different values of
$({m^\prime}^2 \; , \; s_0)$
(Parameter-Set I)}
\label{figure5}
\end{figure}

\begin{figure}
\caption{Variation of $\beta (M^2)$ due to the errorbars in
table\protect\ref{table1} and table\protect\ref{table2} where
${m^\prime}^2 = 1.6 {\rm GeV}^2$ and $s_0 = 1.8 {\rm GeV}^2$
are fixed and taken from fig.\protect\ref{figure4} as the values with
maximal Borel stability in the window $M^2 \ge {M_{min}}^2 = 1.0 {\rm GeV}^2$.}
\label{figure6}
\end{figure}

\begin{figure}
\caption{Borel curve for $\beta$ with 3 different values of $({m^\prime}^2 \; ,
\; s_0)$
(Parameter-Set II)}
\label{figure7}
\end{figure}

\begin{figure}
\caption{Borel curve for ${\xi^\prime}$ with 3 different values of
$({m^\prime}^2 \; , \; s_0)$
(Parameter-Set II)}
\label{figure8}
\end{figure}

\begin{figure}
\caption{The off-shell $\rho$-$\omega$ mixing angle $\theta (q^2)$ with
$\beta =0.78 $ }
\label{figure9}
\end{figure}

\begin{figure}
\caption{The $N-N$ Potential resulting from the $\rho$-$\omega$ mixing
$V_{NN} ^{\rho\omega} (r)$ (in arbitrary units)
calculated with $\lambda =1.78$ (dotted line)
(which is the result for Parameter-Set I (c.f.table\protect\ref{table1}))
in comparison with using the {\em on shell} value for $\theta$
(corresponding to $\lambda =0$)\protect\cite{CB} (full line).
The second figure magnifies the relevant region of $r$, where the sign change
of the off shell curve takes place.}
\label{figure10}
\end{figure}

\begin{figure}
\caption{Isospin breaking nuclear force due to $\rho$ or $\omega$ exchange
with isospin breaking $\rho (\omega)-N-N$ vertex}
\label{figure11}
\end{figure}

\begin{table}
\caption{The sets I-IV of input parameters for the OPE}
\begin{tabular}{ccccc}
       & $\xi \cdot 10^{4} $  &  ${m_d-m_u \over m_d+m_u}$  &
$\gamma \cdot 10^{3} $  &
$\alpha_s \langle \bar{q}q \rangle_0^2 \cdot 10^{4} {\rm GeV}^6$ \\ \tableline
Set\ I   & 11.3 $\pm$ 1.3  &   0.28  $\pm$ 0.03  &   -9.0 $\pm$ 3.0  & 1.81 \\
Set\ II  & 11.3 $\pm$ 1.3  &   0 .28 $\pm$ 0.03  &   -9.0 $\pm$ 3.0  & 3.81 \\
Set\ III & 11.3 $\pm$ 1.3 &   0.28 $\pm$ 0.03    &   -2.0  $\pm$ 1.0& 1.81  \\
Set\ IV  & 11.3 $\pm$ 1.3    &   0.28 $\pm$ 0.03    &   -2.0 $\pm$1.0 & 3.81
\\
\end{tabular}
\label{table1}
\end{table}

\begin{table}
\caption{The OPE coefficients $c_i$ ($i=0\dots3$) calculated with the
parameter sets I-IV from table\protect\ref{table1} }
\begin{tabular}{ccccc}
         & $c_0$  &  $c_1$  &  $c_2$             & $c_3$          \\ \tableline
Set\ I   & 0.147  &   0     &   2.66  $\pm$ 0.30 &- 0.83 $\pm$ 0.23\\
Set\ II  & 0.147  &   0     &   2.66  $\pm$ 0.30 &- 1.75 $\pm$ 0.48\\
Set\ III & 0.147  &   0     &   2.70  $\pm$ 0.30 &- 0.29 $\pm$ 0.08\\
Set\ IV  & 0.147  &   0     &   2.70  $\pm$ 0.30 &- 0.61 $\pm$ 0.17\\
\end{tabular}
\label{table2}
\end{table}

\begin{table}
\caption{$\beta$,${\xi^\prime}$ and ${\beta^\prime}$
from the FESR for different values of
${m^\prime}^2$ and $s_0$ (Parameter-SetI) }
\begin{tabular}{cccc}
$({m^\prime}^2,s_0)\; [{\rm GeV}^2]$  &  $\beta$ & ${\xi^\prime}$ &
${\beta^\prime}$ \\ \tableline
$(1.4\; ,\;2.0)$         &  0.91    & 1.03   & -4.41    \\
$(1.6\; ,\;2.0)$         &  0.75    & 0.87   & -3.79    \\
$(1.8\; ,\;2.0)$         &  0.64    & 0.74   & -3.44    \\
$(2.0\; ,\;2.0)$         &  0.55    & 0.62   & -3.20    \\ \tableline
$(1.8\; ,\;1.4)$         &  0.64    & 0.73   & -3.38    \\
$(1.8\; ,\;2.0)$         &  0.64    & 0.74   & -3.44    \\
$(1.8\; ,\;2.4)$         &  0.64    & 0.73   & -3.52    \\
\end{tabular}
\label{table3}
\end{table}

\begin{table}
\caption{The final values of $\beta$, $\xi^\prime$ and $\beta^\prime$
as well as the resonance mass ${m^\prime}^2$ for
the four different parameter sets of table\protect\ref{table1} extracted from
the Borel stability analysis}
\begin{tabular}{ccccc} & $\beta$ & $\xi^{\prime}$
& $\beta^{\prime}$ & ${m^\prime}^2$ \\ \tableline
Set\ I   & 0.78  $\pm$ 0.07 & 0.98 $\pm$ 0.17 & -3.52 $\pm$ 0.30 & 1.6 \\
Set\ II  & 0.65  $\pm$ 0.13 & 0.85 $\pm$ 0.46 & -2.71 $\pm$ 0.60 & 2.0 \\
Set\ III & 0.47  $\pm$ 0.04 & 0.56 $\pm$ 0.07 & -2.61 $\pm$ 0.15 & 2.4 \\
Set\ IV  & 0.62  $\pm$ 0.06 & 0.76 $\pm$ 0.11 & -3.06 $\pm$ 0.23 & 1.9 \\
\end{tabular}
\label{table4}
\end{table}

\end{document}